\documentclass[longauth]{aa}
\usepackage{graphicx}
\usepackage{txfonts}
\usepackage{xcolor}
\usepackage{hyperref}
\hypersetup{colorlinks, linkcolor=blue, citecolor=blue, urlcolor=blue}

\newcommand{\te}{t_{\rm E}}
\newcommand{\thetae}{\theta_{\rm E}}

\newcommand{\dl}{D_{\rm L}}
\newcommand{\ds}{D_{\rm S}}

\definecolor{brown}{rgb}{0.59, 0.29, 0.0}
\definecolor{darkgreen}{rgb}{0.0, 0.42, 0.24}
\definecolor{darkblue}{rgb}{0.01, 0.31, 0.59}
\definecolor{darkblue}{rgb}{0.0, 0.25, 0.42}
\definecolor{blue}{rgb}{0.0,0.0,1.0}
\definecolor{green}{rgb}{0.0,1.0,0.0}

\begin{document} 

\title{
Six binary brown dwarf candidates identified by microlensing
}
\titlerunning{Microlensing binary brown dwarfs}

\author{
     Cheongho~Han\inst{\ref{cbnu}}
\and Chung-Uk~Lee\inst{\ref{kasi}\thanks{\tt leecu@kasi.re.kr}} 
\and Ian~A.~Bond\inst{\ref{massey}}
\and Andrzej~Udalski\inst{\ref{warsaw}} 
\\
(Leading authors)
\\
     Michael~D.~Albrow\inst{\ref{canterbury}}   
\and Sun-Ju~Chung\inst{\ref{kasi}}      
\and Andrew~Gould\inst{\ref{osu}}      
\and Youn~Kil~Jung\inst{\ref{kasi},\ref{ust}} 
\and Kyu-Ha~Hwang\inst{\ref{kasi}} 
\and Yoon-Hyun~Ryu\inst{\ref{kasi}} 
\and Yossi~Shvartzvald\inst{\ref{weizmann}}   
\and In-Gu~Shin\inst{\ref{cfa}} 
\and Jennifer~C.~Yee\inst{\ref{cfa}}   
\and Weicheng~Zang\inst{\ref{cfa},\ref{tsinghua}}     
\and Hongjing~Yang\inst{\ref{tsinghua}}     
\and Sang-Mok~Cha\inst{\ref{kasi},\ref{kyunghee}} 
\and Doeon~Kim\inst{\ref{cbnu}}
\and Dong-Jin~Kim\inst{\ref{kasi}} 
\and Seung-Lee~Kim\inst{\ref{kasi}} 
\and Dong-Joo~Lee\inst{\ref{kasi}} 
\and Yongseok~Lee\inst{\ref{kasi},\ref{kyunghee}} 
\and Byeong-Gon~Park\inst{\ref{kasi}} 
\and Richard~W.~Pogge\inst{\ref{osu}}
\\
(The KMTNet Collaboration)
\\
     Przemek~Mr{\'o}z\inst{\ref{warsaw}} 
\and Micha{\l}~K.~Szyma{\'n}ski\inst{\ref{warsaw}}
\and Jan~Skowron\inst{\ref{warsaw}}
\and Rados{\l}aw~Poleski\inst{\ref{warsaw}} 
\and Igor~Soszy{\'n}ski\inst{\ref{warsaw}}
\and Pawe{\l}~Pietrukowicz\inst{\ref{warsaw}}
\and Szymon~Koz{\l}owski\inst{\ref{warsaw}} 
\and Krzysztof~A.~Rybicki\inst{\ref{warsaw},\ref{weizmann}}
\and Patryk~Iwanek\inst{\ref{warsaw}}
\and Krzysztof~Ulaczyk\inst{\ref{warwick}}
\and Marcin~Wrona\inst{\ref{warsaw},\ref{villanova}}
\and Mariusz~Gromadzki\inst{\ref{warsaw}}          
\and Mateusz~J.~Mr{\'o}z\inst{\ref{warsaw}} 
\and Micha{\l} Jaroszy{\'n}ski\inst{\ref{warsaw}}
\and Marcin Kiraga\inst{\ref{warsaw}}
\\
(The OGLE Collaboration)
\\
     Fumio~Abe\inst{\ref{nagoya}}
\and David~P.~Bennett\inst{\ref{nasa},\ref{maryland}}
\and Aparna~Bhattacharya\inst{\ref{nasa},\ref{maryland}}
\and Akihiko~Fukui\inst{\ref{tokyo-earth},}\inst{\ref{spain}}
\and Ryusei~Hamada\inst{\ref{osaka}}
\and Stela~Ishitani~Silva\inst{\ref{nasa},\ref{catholic}}  
\and Yuki~Hirao\inst{\ref{tokyo-ast}}
\and Asahi Idei\inst{\ref{osaka}}
\and Shota~Miyazaki\inst{\ref{jaxa}}
\and Yasushi~Muraki\inst{\ref{nagoya}}
\and Tutumi~Nagai\inst{\ref{osaka}}
\and Kansuke~Nunota\inst{\ref{osaka}}
\and Greg~Olmschenk\inst{\ref{nasa}}
\and Cl{\'e}ment~Ranc\inst{\ref{sorbonne}}
\and Nicholas~J.~Rattenbury\inst{\ref{auckland}}
\and Yuki~Satoh\inst{\ref{yokohama}}
\and Takahiro~Sumi\inst{\ref{osaka}}
\and Daisuke~Suzuki\inst{\ref{osaka}}
\and Takuto Tamaoki\inst{\ref{osaka}}
\and Sean K. Terry\inst{\ref{nasa}, \ref{maryland}}
\and Paul~J.~Tristram\inst{\ref{john}}
\and Aikaterini~Vandorou\inst{\ref{nasa},\ref{maryland}}
\and Hibiki~Yama\inst{\ref{osaka}}
\\
(The MOA Collaboration)
}

\institute{
      Department of Physics, Chungbuk National University, Cheongju 28644, Republic of Korea                                                          \label{cbnu}     
\and  Korea Astronomy and Space Science Institute, Daejon 34055, Republic of Korea                                                                    \label{kasi}   
\and  Institute of Natural and Mathematical Science, Massey University, Auckland 0745, New Zealand                                                    \label{massey}    
\and  Astronomical Observatory, University of Warsaw, Al.~Ujazdowskie 4, 00-478 Warszawa, Poland                                                      \label{warsaw}   
\and  University of Canterbury, Department of Physics and Astronomy, Private Bag 4800, Christchurch 8020, New Zealand                                 \label{canterbury}  
\and  Department of Astronomy, Ohio State University, 140 West 18th Ave., Columbus, OH 43210, USA                                                     \label{osu} 
\and  University of Science and Technology, Daejeon 34113, Republic of Korea                                                                          \label{ust}
\and  Department of Particle Physics and Astrophysics, Weizmann Institute of Science, Rehovot 76100, Israel                                           \label{weizmann}   
\and  Center for Astrophysics $|$ Harvard \& Smithsonian 60 Garden St., Cambridge, MA 02138, USA                                                      \label{cfa}  
\and  Department of Astronomy and Tsinghua Centre for Astrophysics, Tsinghua University, Beijing 100084, China                                        \label{tsinghua} 
\and  School of Space Research, Kyung Hee University, Yongin, Kyeonggi 17104, Republic of Korea                                                       \label{kyunghee}     
\and  Department of Physics, University of Warwick, Gibbet Hill Road, Coventry, CV4 7AL, UK                                                           \label{warwick}
\and  Villanova University, Department of Astrophysics and Planetary Sciences, 800 Lancaster Ave., Villanova, PA 19085, USA                           \label{villanova} 
\and  Institute for Space-Earth Environmental Research, Nagoya University, Nagoya 464-8601, Japan                                                     \label{nagoya}     
\and  Code 667, NASA Goddard Space Flight Center, Greenbelt, MD 20771, USA                                                                            \label{nasa} 
\and  Department of Astronomy, University of Maryland, College Park, MD 20742, USA                                                                    \label{maryland}  
\and  Department of Earth and Planetary Science, Graduate School of Science, The University of Tokyo, 7-3-1 Hongo, Bunkyo-ku, Tokyo 113-0033, Japan   \label{tokyo-earth} 
\and  Instituto de Astrof{\'i}sica de Canarias, V{\'i}a L{\'a}ctea s/n, E-38205 La Laguna, Tenerife, Spain                                            \label{spain} 
\and  Department of Earth and Space Science, Graduate School of Science, Osaka University, Toyonaka, Osaka 560-0043, Japan                            \label{osaka}  
\and  Department of Physics, The Catholic University of America, Washington, DC 20064, USA                                                            \label{catholic} 
\and  Institute of Astronomy, Graduate School of Science, The University of Tokyo, 2-21-1 Osawa, Mitaka, Tokyo 181-0015, Japan                        \label{tokyo-ast}
\and  Institute of Space and Astronautical Science, Japan Aerospace Exploration Agency, 3-1-1 Yoshinodai, Chuo, Sagamihara, Kanagawa 252-5210, Japan  \label{jaxa}
\and  Sorbonne Universit\'e, CNRS, UMR 7095, Institut d'Astrophysique de Paris, 98 bis bd Arago, 75014 Paris, France                                  \label{sorbonne}
\and  Department of Physics, University of Auckland, Private Bag 92019, Auckland, New Zealand                                                         \label{auckland}    
\and  College of Science and Engineering, Kanto Gakuin University, Yokohama, Kanagawa 236-8501, Japan                                                 \label{yokohama}
\and  University of Canterbury Mt.~John Observatory, P.O. Box 56, Lake Tekapo 8770, New Zealand                                                       \label{john}  
}                                                                                                                                                       
\date{Received ; accepted}

\abstract
{}
{
In single-lens microlensing events, the event timescale ($t_{\rm E}$) is typically 
the only measurable parameter that constrains the lens mass. Since $t_{\rm E}$ 
scales with the square root of the lens mass ($t_{\rm E} \propto M^{1/2}$), a short 
duration may suggest a low-mass lens, such as a brown dwarf (BD). However, a short 
$t_{\rm E}$ can also result from a high relative proper motion between the lens 
and the source, making it difficult to uniquely identify BD candidates based on 
timescale alone. In contrast, binary-lens events often allow for the measurement 
of the angular Einstein radius ($\theta_{\rm E}$) in addition to $t_{\rm E}$. When 
both $t_{\rm E}$ and $\theta_{\rm E}$ are small, the likelihood that the lens is 
of low mass increases significantly. In this study, we analyze microlensing events 
from the 2023 and 2024 observing seasons to identify cases likely caused by binary 
systems composed of BDs.
}
{
By applying criteria that the binary-lens events exhibit well-resolved caustics, 
short time scales ($t_{\rm E} \lesssim 9$ days), and have small angular Einstein 
radii ($\theta_{\rm E} \lesssim 0.17$~mas), we identify six candidate binary BD events: 
MOA-2023-BLG-331, KMT-2023-BLG-2019, KMT-2024-BLG-1005, KMT-2024-BLG-1518, 
MOA-2024-BLG-181, and KMT-2024-BLG-2486. Analysis of these events leads to models 
that provide precise estimates for both lensing observables, $t_{\rm E}$ and 
$\theta_{\rm E}$.
}
{
We estimate the masses of the binary components through Bayesian analysis, utilizing 
the constraints from $\te$ and $\thetae$. The results show that for the events 
KMT-2024-BLG-1005, KMT-2024-BLG-1518, MOA-2024-BLG-181, and KMT-2024-BLG-2486, the 
probability that both binary components lie within the BD mass range exceeds 50\%, 
indicating a high likelihood that the lenses of these events are binary BDs. In 
contrast, for MOA-2023-BLG-331L and KMT-2023-BLG-2019L, the probabilities that the 
lower-mass components of the binary lenses lie within the BD mass range exceed 50\%, 
while the probabilities for the heavier components are below 50\%, suggesting that 
these systems are more likely to consist of a low-mass M dwarf and a BD.  The 
brown-dwarf nature of the binary candidates can ultimately be confirmed by combining 
the measured lens-source relative proper motions with high-resolution imaging taken 
at a later time.
}
{}


\maketitle

\section{Introduction} \label{sec:one}

In microlensing, the event time scale ($\te$) and angular Einstein radius ($\thetae$) 
are key observables for constraining the lens mass $M$, as they are related to the 
mass through the relations
\begin{equation}
\te = {\thetae \over \mu};   \qquad
\thetae = \sqrt{\kappa M \pi_{\rm rel}}.
\label{eq1}
\end{equation}
Here $\kappa = 4G/(c^2 {\rm AU}) \simeq 8.14~{\rm mas}/M_\odot$, $\mu$ represents 
the relative lens-source proper motion, $\pi_{\rm rel} = {\rm AU} (1/\dl - 1/\ds)$ 
is the relative parallax between the lens and source, and $\dl$ and $\ds$ denote 
the distances to the lens source, respectively \citep{Gould2000a}.  For Galactic 
lensing events, these observables scale with the lens mass as:
\begin{equation}
\te \simeq 10~{\rm days}~\left( { M\over 0.2~M_\odot}\right)^{1/2}; \ \ \ 
\thetae \simeq 0.2~{\rm mas}~\left( { M\over 0.2~M_\odot}\right)^{1/2}.
\label{eq2}
\end{equation}
Thus, for events characterized by a short time scale with $\te \lesssim 6.3$~days and 
a small angular Einstein radius with $\theta_{\rm E} \lesssim 0.12$~mas, the lens is 
likely to be a brown dwarf (BD) with a mass below the hydrogen-burning limit.

\begin{table*}[t]
\caption{Event correspondence, alert dates, and coordinates.  \label{table:one}}
\begin{tabular}{llllllllllcc}
\hline\hline
\multicolumn{3}{c}{Event (Alert date)}              &
\multicolumn{1}{c}{$({\rm RA}, {\rm DEC})_{2000}$}  &
\multicolumn{1}{c}{$(l, b)$}                        \\
\multicolumn{1}{c}{KMTNet}                          &
\multicolumn{1}{c}{OGLE}                            &
\multicolumn{1}{c}{MOA}                             &
\multicolumn{1}{c}{}                                &
\multicolumn{1}{c}{}                               \\
\hline
 KMT-2023-BLG-1819   &  OGLE-2023-BLG-0970   &  MOA-2023-BLG-331    &  (18:03:53.20, -27:39:08.10)      &  (3$^\circ\hskip-2pt$.1400, -2$^\circ\hskip-2pt$.8218)   \\
 (2023-07-25)        &  (2023-07-25)         &  (2023-07-24)        &                                   &                                                          \\
\hline
 KMT-2023-BLG-2019   &                       &                      &  (17:50:49.86, -30:48:11.92)      &  (1$^\circ\hskip-2pt$.0163, -1$^\circ\hskip-2pt$.9312)   \\
 (2023-08-17)        &                       &                      &                                   &                                                          \\
\hline
 KMT-2024-BLG-1005   &  OGLE-2024-BLG-0628   &                      &   (17:59:13.34, -28:45:20.81)     &  (1$^\circ\hskip-2pt$.6716, -2$^\circ\hskip-2pt$.4721)   \\
 (2024-05-16)        &  (2024-06-02)         &                      &                                   &                                                          \\
\hline
 KMT-2024-BLG-1518   &  OGLE-2024-BLG-0825   &                      &  (18:01:37.17, -28:57:39.10)      &  (1$^\circ\hskip-2pt$.7532, -3$^\circ\hskip-2pt$.0298)   \\
 (2024-06-25)        &  (2024-06-29)         &                      &                                   &                                                          \\
\hline
 KMT-2024-BLG-2185   &  OGLE-2024-BLG-1086   &  MOA-2024-BLG-181    &  (17:58:11.33, -29:14:42.90)      &  (1$^\circ\hskip-2pt$.1341, -2$^\circ\hskip-2pt$.5198)   \\
 (2024-08-12)        &  (2024-08-16)         &  (2024-08-09)        &                                   &                                                          \\
\hline
 KMT-2024-BLG-2486   &  OGLE-2024-BLG-1199   &                      &  (17:56:37.05, -28:56:11.18)      &  (1$^\circ\hskip-2pt$.2300, -2$^\circ\hskip-2pt$.0681)   \\
 (2024-09-09)        &  (2024-09-12)         &                      &                                   &                                                          \\
\hline
\end{tabular}
\end{table*}

In single-lens events, the event time scale is usually the only observable that can be 
derived from the light curve analysis. In such cases, a short time scale may be attributed 
to a high relative proper motion between the lens and source, rather than indicating 
a low-mass lens.  Consequently, it is difficult to conclusively infer that the lens mass 
lies in the substellar regime based solely on $\te$. Although microlensing is a powerful 
tool for detecting BDs, it remains challenging to identify reliable BD candidates from 
single-lens events.  \footnote{ Under specific observational conditions, 
when a short timescale, high-magnification event is observed simultaneously from two sites, 
the lens mass can be uniquely determined by measuring the microlens parallax ($\pi_{\rm E}$) 
\citep{Refsdal1966, Gould1994}. This method was employed to identify three isolated brown 
dwarfs, OGLE-2007-BLG-224 \citep{Gould2009}, OGLE-2017-BLG-0896L \citep{Shvartzvald2019}, 
and OGLE-2017-BLG-1186L \citep{Li2019}, through the measurement of $\pi_{\rm E}$.  }

In binary-lens events involving two lens components, the probability of measuring the 
additional observable $\thetae$ is high, as a significant fraction these events exhibit 
caustic-crossing features in their light curves.  When both observables, $\te$ and $\thetae$, 
are measured and found to be small, the likelihood that the lens has a low mass increases 
significantly. The event time scale and angular Einstein radius corresponding to each lens 
component is related to the mass ratio $q$ between the components by the following relations:
\begin{equation}
t_{{\rm E},1} = \sqrt{{1\over 1+q}} t_{\rm E}; \qquad 
t_{{\rm E},2} = \sqrt{{q\over 1+q}} t_{\rm E} 
\label{eq3}
\end{equation}
for the event time scale and
\begin{equation}
\theta_{{\rm E},1} = \sqrt{{1\over 1+q}} \thetae; \qquad 
\theta_{{\rm E},2} = \sqrt{{q\over 1+q}} \theta_{\rm E}  
\label{eq4}
\end{equation}
for the angular Einstein radius.
Here $t_{\rm E}$ and $\theta_{\rm E}$ are the values corresponding to the total mass of 
the binary lens system. If the binary lens consists of roughly equal-mass components, 
then events with $\te \lesssim 9$ days and $\thetae \lesssim 0.17$ mas suggest that
each component is likely a BD with substellar mass.

An alternative approach to identifying BD events in binary-lens systems focuses on 
microlensing events with low mass ratios between the lens components. This method is 
motivated by the fact that typical Galactic microlensing events are primarily caused 
by low-mass stars \citep{Han2003}, and companions with small mass ratios are likely to 
be BDs.  By applying a selection criterion of $q \lesssim 0.1$ to binary-lens 
events identified in microlensing survey data, dozens of BD candidates have been 
reported in previous studies \citep{Han2022, Han2023a, Han2023b, Han2024}.  In 
these cases, only one component of the binary lens is inferred to be a BD.

BDs are often described as "failed stars," because they are too massive 
to be considered planets but not massive enough to sustain hydrogen fusion like 
ordinary stars. Because their physical properties place them in this in-between 
regime, their formation pathways remain uncertain.

The identification of binary BDs is valuable because the frequency, separation, and 
mass ratios of such systems provide important clues regarding whether they 
form in a stellar-like manner, through cloud fragmentation \citep{Bate2002}, or in 
a planet-like manner, through disk instability or core accretion \citep{Whitworth2005}.

The microlensing method is particularly powerful in this context because it can 
reveal systems that are extremely faint, cold, and inaccessible to direct imaging.
Moreover, microlensing surveys annually detect hundreds of lensing events caused 
by binary systems, with a significant fraction arising from binaries composed of 
BDs.  The resulting datasets can provide valuable constraints on how common such 
binaries are in our galaxy and help in testing competing formation theories.

In this study, we examine microlensing data from the 2023 and 2024 observing seasons 
to search for events generated by binary systems composed of BDs. Our selection 
criteria require that both key lensing parameters, $\te$ and $\thetae$, are securely 
measured and lie below defined threshold values. Based on these criteria, we identify 
six candidate binary-lens events in which both components are likely BDs.

\section{Selection of candidate events and data} \label{sec:two}

To identify binary BD candidates, we first examined binary-lens events detected from 
the Korea Microlensing Telescope Network \citep[KMTNet;][]{Kim2016} survey during the 
2023 and 2024 seasons. The primary selection criterion was that the event exhibits 
well-resolved caustic features, ensuring that the angular Einstein radius can be 
securely measured. Additionally, we applied the criteria of $\te \lesssim 9$ days and 
$\thetae \lesssim 0.17$ mas. Using these criteria, we identified six candidate events: 
KMT-2023-BLG-1819, KMT-2023-BLG-2019, KMT-2024-BLG-1005, KMT-2024-BLG-1518, 
KMT-2024-BLG-2185, and KMT-2024-BLG-2486.

All of these events, with the exception of KMT-2023-BLG-2019, were also observed by 
other lensing surveys conducted by the Optical Gravitational Lensing Experiment 
\citep[OGLE;][]{Udalski2015} and the Microlensing Observations in Astrophysics 
\citep[MOA;][]{Bond2001, Sumi2003} collaborations.  In Table~\ref{table:one}, we 
provide a summary of the event correspondences, along with the Equatorial and Galactic 
coordinates for each event, as well as the dates when alerts were issued by the 
individual survey groups. Among the events, four (KMT-2023-BLG-2019, KMT-2024-BLG-1005, 
KMT-2024-BLG-1518, and KMT-2024-BLG-2486) were initially identified by the KMTNet group, 
while two (MOA-2023-BLG-331 and MOA-2024-BLG-181) were first detected by the MOA group.  
For events observed by multiple surveys, we use the event ID assigned by the first 
detecting group as the representative designation, following the convention in the 
microlensing community.

Photometric data for the events were obtained from observations conducted by the individual
surveys. The KMTNet survey employs three identical telescopes, strategically located in the
Southern Hemisphere for continuous coverage of lensing events: at Siding Spring Observatory 
in Australia (KMTA), Cerro Tololo Interamerican Observatory in Chile (KMTC), and South 
African Astronomical Observatory in South Africa (KMTS). Each KMTNet telescope has a 
1.6-meter aperture and is equipped with a camera that provides a 4-square-degree field of 
view. The MOA survey uses a 1.8-meter telescope situated at Mt. John University Observatory 
in New Zealand, with a camera that covers a 2.2-square-degree field of view. The OGLE survey 
is conducted using the 1.3-meter Warsaw Telescope, located at Las Campanas Observatory in 
Chile. The camera mounted on the OGLE telescope has a field of view of 1.4 square degrees. 
Observations from the KMTNet and OGLE surveys were primarily carried out in the $I$ band, 
whereas the MOA survey observations were made in a customized MOA-$R$ band, with a 
wavelength range of 609 to 1109 nm.

The data used in the analyses were obtained by performing 
photometry with the pipelines developed by the respective survey groups: \citet{Albrow2009} 
for KMTNet, \citet{Bond2001} for MOA, and \citet{Udalski2003} for OGLE.
To correct the photometric uncertainties provided by the automated pipeline, 
we applied a rescaling procedure. The goal was to bring the reported error bars 
into agreement with the actual scatter in the measurements, and to adjust them 
such that the reduced $\chi^2$ of the model evaluation approached unity for each 
individual dataset. This adjustment followed the approach described by \citet{Yee2012}.

In the case of KMTNet data, short-duration correlated noise may arise from residual 
images of previous exposures. Although microlensing light curves may exhibit diverse 
distortions, they rarely reproduce noise patterns found in observational data. On the 
rare occasions when a model reproduces very short-duration features, we inspected the 
images at the corresponding epochs to assess whether the signal is genuine. For the 
analyzed events, no such cases were identified.

\section{Modeling light curves} \label{sec:three}

Given the caustic-crossing features in all the analyzed events, we modeled the light 
curves using the binary-lens single-source (2L1S) configuration. In this model, the 
light curve is described by seven parameters. Three of these parameters $(t_0, u_0, \te)$ 
represent the lens-source approach, with $t_0$ denoting the time of closest approach, 
$u_0$ indicating the separation at that time, and $\te$ representing the event time scale. 
Two additional parameters $(s, q)$ characterize the binary lens, where $s$ is the projected 
separation between the lens components ($M_1$ and $M_2$) and $q$ is their mass ratio. The 
parameters $u_0$ and $s$ are scaled to the angular Einstein radius corresponding to the 
total mass of the binary lens. The parameter $\alpha$ defines the direction of the source's 
motion relative to the $M_1$--$M_2$ axis. The final parameter, $\rho$, represents the ratio 
of the angular source radius ($\theta_*$) to the angular Einstein radius, $\rho = \theta_* 
/ \thetae$. This parameter (normalized source radius) accounts for finite-source effects, 
particularly in the caustic-crossing regions. Since all the lensing events have short 
durations, we do not consider long-term higher-order effects, such as the parallax effect 
from Earth's orbital motion around the Sun \citep{Gould1992, Gould2000a} or the lens-orbital 
effect from the binary lens \citep{Alcock2000, Bennett1999, Albrow2000}.

\begin{table*}[t]
\caption{Lensing solutions of MOA-2023-BLG-331, KMT-2023-BLG-2019, and KMT-2024-BLG-1005.\label{table:two}}
\begin{tabular}{lllllll}
\hline\hline
\multicolumn{1}{c}{Parameter}            &
\multicolumn{1}{c}{MOA-2023-BLG-331}     &
\multicolumn{1}{c}{KMT-2023-BLG-2019}    &
\multicolumn{1}{c}{KMT-2024-BLG-1005}    \\
\hline
 $t_0$ (HJD$^\prime$)      &  $150.7384 \pm 0.0010 $   &  $170.779 \pm 0.011$     &  $447.513 \pm 0.010$    \\
 $u_0$                     &  $0.08949 \pm 0.00092 $   &  $-0.03 \pm 0.20   $     &  $0.1018 \pm 0.0042$    \\
 $\te$ (days)              &  $6.418 \pm 0.050     $   &  $6.328 \pm 0.088  $     &  $3.593 \pm 0.021  $    \\
 $s$                       &  $1.3308 \pm 0.0060   $   &  $1.5312 \pm 0.0061$     &  $0.9494 \pm 0.0031$    \\
 $q$                       &  $1.545 \pm 0.0088    $   &  $0.334 \pm 0.019  $     &  $0.362 \pm 0.011  $    \\
 $\alpha$ (rad)            &  $1.52902 \pm 0.00039 $   &  $3.612 \pm 0.010  $     &  $4.176 \pm 0.013  $    \\
 $\rho$ ($10^{-3}$)        &  $5.894 \pm 0.051     $   &  $11.6 \pm 0.22    $     &  $9.32 \pm 0.17    $    \\
\hline
\end{tabular}
\tablefoot{ ${\rm HJD}^\prime = {\rm HJD}- 2460000$.  }
\end{table*}

In the modeling process, we searched for the set of lensing parameters (lensing solution) 
that best reproduces the observed light curve. In the initial stage, we grouped the parameters 
into two categories: $(s, q)$ in one group and the remaining parameters in the other. We 
performed a grid search over the $(s, q)$ parameter space, while the remaining parameters 
were optimized using a downhill method based on the Markov Chain Monte Carlo (MCMC) technique. 
This first modeling stage yielded a $\chi^2$ map across the grid space, from which we 
identified local solutions. In the second stage, each of these local solutions was refined 
by allowing all parameters to vary freely.  In cases for which multiple solutions provided 
comparably good fits to the data, we present all degenerate solutions. However, for all 
events analyzed in this study, a unique solution was found. In the following subsections, 
we detail the analysis for each individual event.

\begin{figure}[t]
\includegraphics[width=\columnwidth]{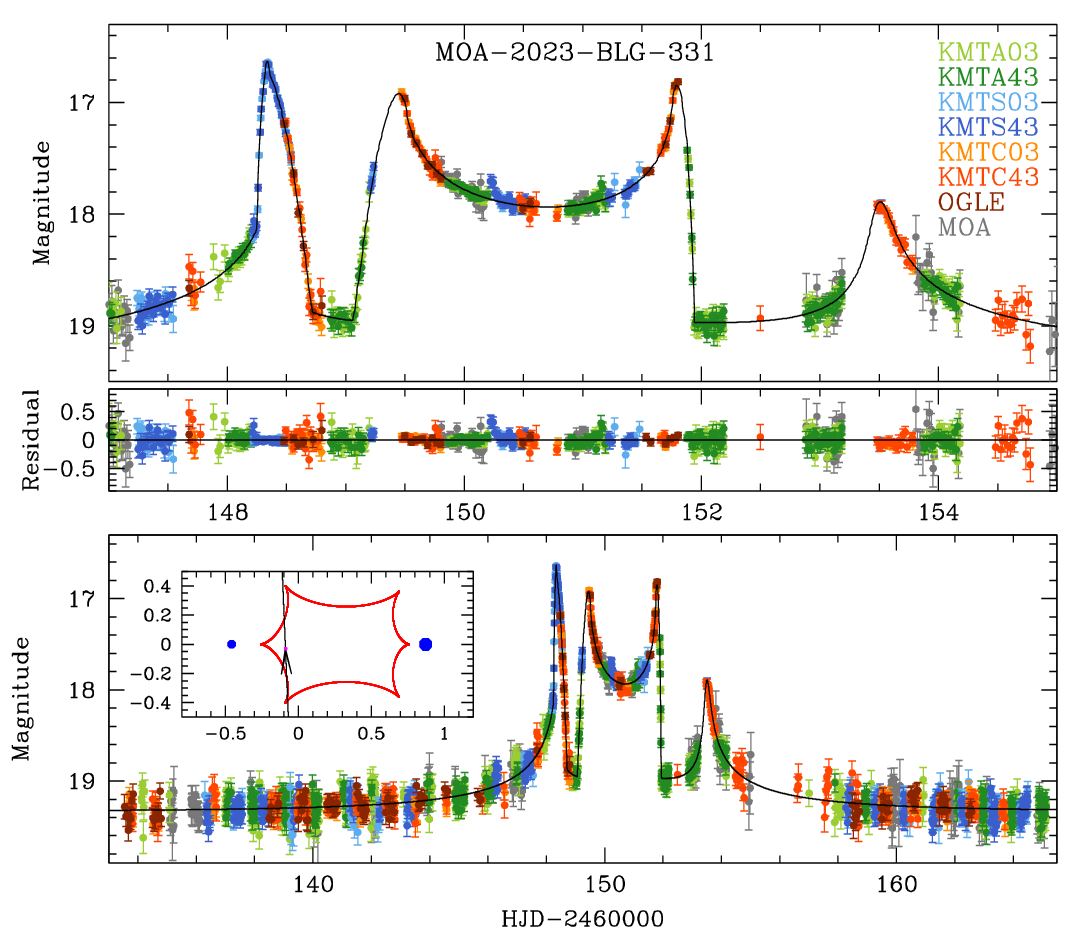}
\caption{
Light curve of the lensing event MOA-2023-BLG-331.  The bottom panel displays the full 
light curve, while the upper panels provide a zoomed-in view around the anomaly features 
and the residuals from the model.  The curve overlaid on the data points represents the 
best-fit model.  The inset in the bottom panel illustrates the lens system configuration, 
with the red cusped shape representing the caustic, the two blue dots indicating the 
positions of the lens components (with the larger dot representing the heavier lens 
component), and the arrowed line depicting the source's trajectory.
}
\label{fig:one}
\end{figure}

\subsection{MOA-2023-BLG-331} \label{sec:three-one}

The lensing event MOA-2023-BLG-331 was first detected by the MOA survey on July 24, 2023,
corresponding to the abridged Heliocentric Julian date ${\rm HJD}^\prime \equiv {\rm HJD} -
2460000 =150$, and was later identified by the KMTNet and OGLE surveys the following day.
The source is located in the overlapping region between the KMTNet prime fields BLG03 and
BLG43, with observations conducted at a 30-minute cadence for each field, and a 15-minute
cadence for combined observations. The baseline $I$-band magnitude of the source was 
$I_{\rm base} = 20.03$, and the extinction toward the field was $A_I = 1.05$.

The lensing light curve for MOA-2023-BLG-331 is shown in Figure~\ref{fig:one}. Although 
the primary magnification occurred within a span of less than 10 days, the light curve 
was well-covered by dense data, revealing intricate anomaly features. The two spikes at 
${\rm HJD}^\prime \sim 149.4$ and 151.8, along with the U-shaped region between them, 
indicate that these features are due to caustic crossings of the source. The bumps around 
${\rm HJD}^\prime \sim 148.3$ and 153.5 appear to be caused by the source's approach to 
the cusps of a binary caustic.

From a 2L1S modeling, we found a unique solution that explains all the anomalous features. 
The model curve is overlaid on the data points in Figure~\ref{fig:one}, and the corresponding 
model parameters are summarized in Table~\ref{table:two}.  The binary lens parameters, $(s, 
q) \sim (1.33, 1.55)$, suggest that the lens is a binary system with components of similar 
masses and a projected separation slightly larger than the Einstein radius.  We note that 
a mass ratio greater than unity indicates that the source trajectory lay closer to the 
lower-mass component of the binary lens system. The estimated event time scale, $\te = 
(6.418 \pm 0.050)$ days, is relatively short.  The time scales corresponding to the 
individual lens components are $t_{{\rm E},1} \sim 4.0$ days and $t_{{\rm E},2} \sim 5.0$ 
days, suggesting that both components are of low mass.  The normalized source radius, 
$\rho = (5.894 \pm 0.051) \times 10^{-3}$, is measured with high precision due to the 
dense coverage of the caustic crossings.  As will be discussed in Section~\ref{sec:four}, 
the source is a main-sequence star. In lensing events involving an M-dwarf lens and a 
main-sequence source, the normalized source radius is typically $\rho \lesssim 2 \times 
10^{-3}$. Given that the angular Einstein radius is determined by $\thetae = \theta_*/
\rho$, where $\theta_*$ is the angular source radius, a measured $\rho$ value significantly 
larger than the typical range implies that the angular Einstein radius is likely small.

The inset in the bottom panel of Figure~\ref{fig:one} illustrates the configuration of the 
lens system. The binary lens created a resonant caustic structure featuring six cusps. The 
source star passed through the left side of the caustic, near the lower component of the 
lens, at an angle close to perpendicular. It first crossed the tip of the lower-left cusp, 
producing the initial bump in the light curve. It then entered and exited the left section 
of the caustic, resulting in a pair of sharp caustic spikes. After exiting, the source 
approached the upper-left cusp, giving rise to the final bump.

\subsection{KMT-2023-BLG-2019} \label{sec:three-two}

The lensing event KMT-2023-BLG-2019 was observed exclusively by the KMTNet survey. The 
source star lies in the overlapping region of two KMTNet prime fields, BLG01 and BLG41, 
both monitored at a high cadence of 30 minutes, providing dense temporal coverage of the 
light curve. The source had a baseline $I$-band magnitude of $I_{\rm base} = 20.4$, with 
an extinction of $A_I = 2.65$ toward the field. The event was first identified on August 
17, 2023 (corresponding to ${\rm HJD}^\prime = 173$), after the light curve exhibited 
clear anomalies caused by caustic crossings.

\begin{figure}[t]
\includegraphics[width=\columnwidth]{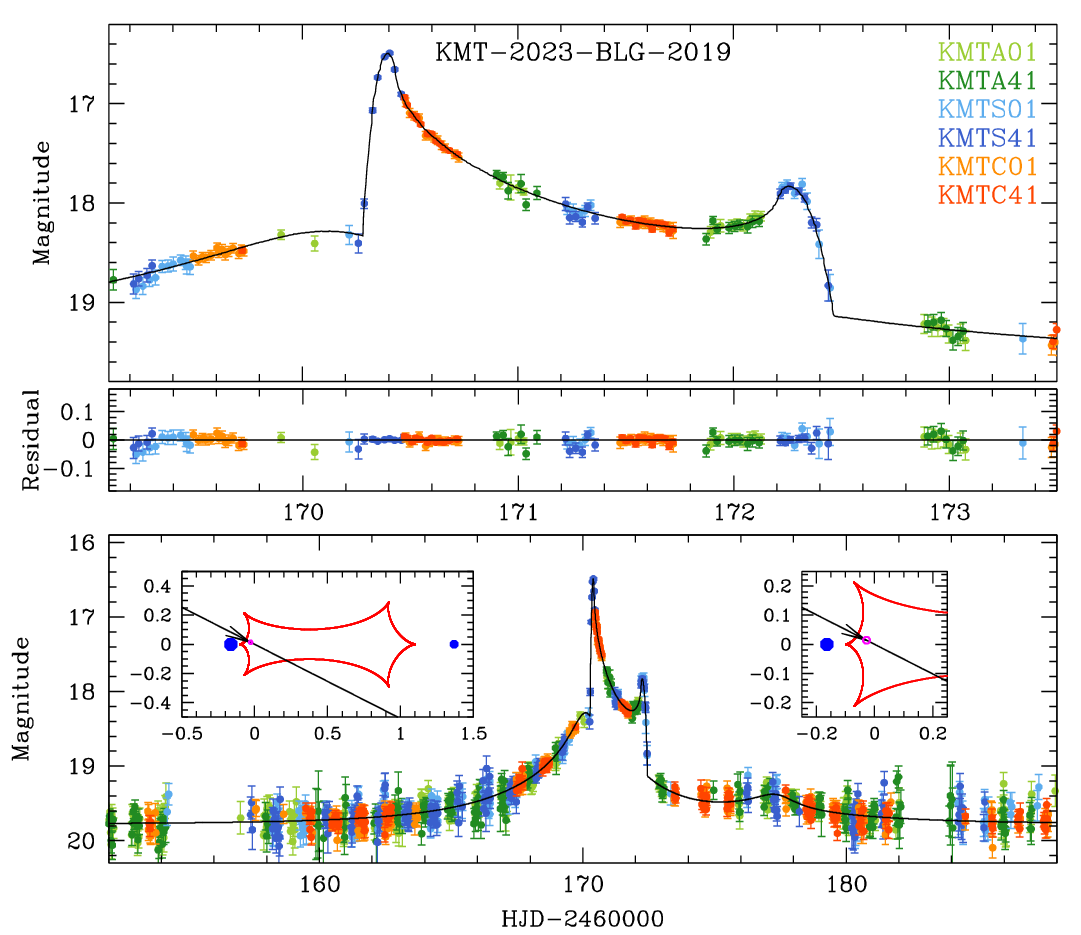}
\caption{
Lensing light curve of KMT-2023-BLG-2019. The notations are the same as those used in 
Fig.~\ref{fig:one}. In the bottom panel, the left inset displays the full view of the 
caustic, while the right inset provides a close-up of the region where the source enters 
the caustic.
}
\label{fig:two}
\end{figure}

Figure~\ref{fig:two} presents the lensing light curve of KMT-2023-BLG-2019, which is 
marked by a pair of prominent caustic-crossing spikes occurring at ${\rm HJD}^\prime 
\sim 170.4$ and $\sim 172.3$. Both caustic features were densely resolved, with each 
spike covered by the KMTS observations. In addition to these features, the light curve 
exhibits a weak bump around ${\rm HJD}^\prime \sim 177$.  As with the previous event, 
the overall duration of the event is short, with the primary magnification phase 
concluding in under 10 days.

A 2L1S modeling provides a unique solution that accounts for all the observed anomalous 
features. In Figure~\ref{fig:two}, we show the model curve superimposed on the data points. 
The complete set of lensing parameters is listed in Table~\ref{table:two}. The derived 
binary parameters, $(s, q) \sim (1.53, 0.33)$, indicate that the lens is a binary system 
with components of comparable masses, separated by slightly more than the angular Einstein 
radius. As expected from the brief lensing magnification episode, the event time scale is 
relatively short, with $\te = (6.328 \pm 0.088)$ days. The individual time scales for the 
lens components are $t_{{\rm E},1} \sim 5.4$~days and $t_{{\rm E},2} \sim 3.1$~days.  The 
well-resolved caustic features allowed for a precise measurement of the normalized source 
radius, $\rho = (11.6 \pm 0.22) \times 10^{-3}$. Given that the source is a main-sequence 
star, the unusually large value of $\rho$ implies a small angular Einstein radius. Together 
with the short event time scales, this suggests that the lens has a low mass.

The configuration of the lens system is shown in the insets of the bottom panel, revealing 
a six-sided resonant caustic elongated along the binary axis. The source trajectory passed 
near the more massive lens component, first crossing the upper-left fold and then the lower 
fold of the caustic. These crossings account for the observed caustic-crossing features in 
the light curve. After exiting the caustic, the source passed near the lower-left cusp, 
which produced the weak bump observed in the light curve.

\begin{table*}[t]
\caption{Lensing solutions of KMT-2024-BLG-1518, MOA-2024-BLG-181, and KMT-2024-BLG-2486.\label{table:three}}
\begin{tabular}{lllllll}
\hline\hline
\multicolumn{1}{c}{Parameter}            &
\multicolumn{1}{c}{KMT-2024-BLG-1518}     &
\multicolumn{1}{c}{MOA-2024-BLG-181}    &
\multicolumn{1}{c}{KMT-2024-BLG-2486}    \\
\hline
 $t_0$ (HJD$^\prime$)  &   $485.255 \pm 0.019$    &  $531.8644 \pm 0.0041$    &  $561.3412 \pm 0.0015$     \\
 $u_0$                 &   $0.1297 \pm 0.0020$    &  $5.974 \pm 0.056    $    &  $2.581 \pm 0.046    $     \\
 $\te$ (days)          &   $5.118 \pm 0.076  $    &  $6.660 \pm 0.049    $    &  $7.395 \pm 0.047    $     \\
 $s$                   &   $0.8242 \pm 0.0056$    &  $2.185 \pm 0.010    $    &  $3.080 \pm 0.015    $     \\
 $q$                   &   $0.901 \pm 0.029  $    &  $1.722 \pm 0.052    $    &  $1.021 \pm 0.064    $     \\
 $\alpha$ (rad)        &   $2.235 \pm 0.016  $    &  $0.3593 \pm 0.0030  $    &  $0.1091 \pm 0.0043  $     \\
 $\rho$ ($10^{-3}$)    &   $8.60 \pm 0.11    $    &  $6.16 \pm 0.12      $    &  $5.231 \pm 0.096    $     \\
\hline
\end{tabular}
\end{table*}

\begin{figure}[t]
\includegraphics[width=\columnwidth]{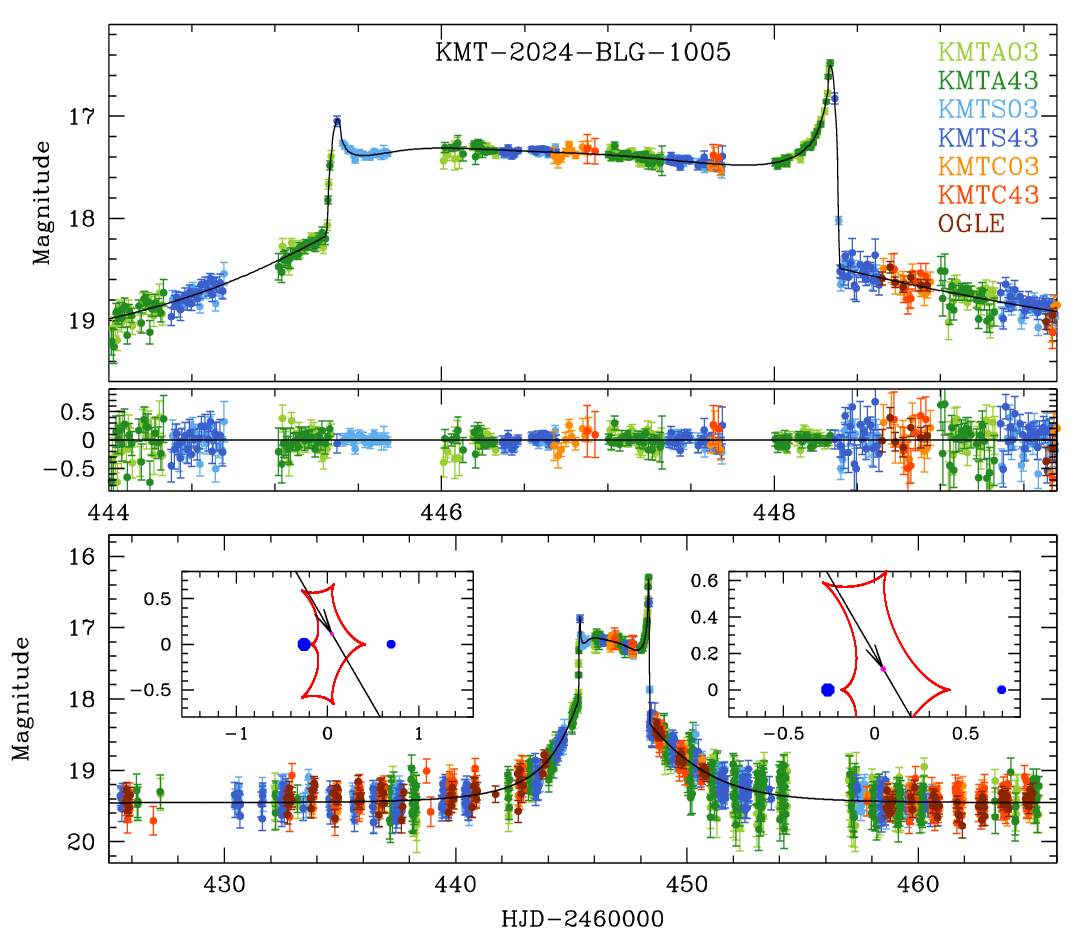}
\caption{
Lensing light curve and lens-system configuration of KMT-2024-BLG-1005.
}
\label{fig:three}
\end{figure}

\subsection{KMT-2024-BLG-1005} \label{sec:three-three}

The lensing event KMT-2024-BLG-1005 involved a source with a baseline magnitude of 
$I_{\rm base} = 19.2$. It was first detected by the KMTNet group on May 16, 2024, and 
later confirmed by the OGLE group. The source is located within the region covered by 
the KMTNet prime fields BLG03 and BLG43, which were observed with a combined cadence 
of 15 minutes. The extinction in this field is $A_I = 1.24$.

Figure~\ref{fig:three} presents the light curve of the event, which, like previous 
cases, is characterized by a short duration, with the main magnification phase lasting 
less than 10 days. The light curve features two distinct and well-defined caustic 
spikes occurring at approximately ${\rm HJD}^\prime \sim 445.3$ and ${\rm HJD}^\prime 
\sim 448.3$.  Notably, the region between these two caustic spikes does not follow the 
typical U-shaped pattern commonly seen in caustic-crossing binary-lens events. Instead, 
it displays a relatively flat, plateau-like structure, indicating a deviation from the 
standard morphology.  Both caustic features were clearly resolved, owing to the dense 
and continuous observational coverage provided by the combined data from the KMTA and 
KMTS observatories.

We modeled the light curve using a 2L1S configuration and found a unique solution 
that explains the anomalous features. The binary-lens parameters, $(s, q) \sim (0.95, 
0.36)$, suggest that the lens consists of two objects with comparable masses and a 
projected separation near the Einstein radius. The event time scale is $\te = (3.593 
\pm 0.021)$ days, which is even shorter than those of the previous two events. The 
time scales corresponding to the individual lens components are $t_{{\rm E},1} \sim 3.1$ 
days and $t_{{\rm E},2} \sim 1.8$ days. From the analysis of the resolved caustics, we 
determined the normalized source radius to be $\rho = (9.32 \pm 0.17) \times 10^{-3}$. 
Given that the source is a main-sequence star, this relatively large $\rho$ value 
suggests a small angular Einstein radius. Combined with the short event time scale, 
this implies that the lens is likely a low-mass object. The complete set of lensing 
parameters is provided in Table~\ref{table:two}, and the corresponding model curve is 
shown in Figure~\ref{fig:three}.

\begin{figure}[t]
\includegraphics[width=\columnwidth]{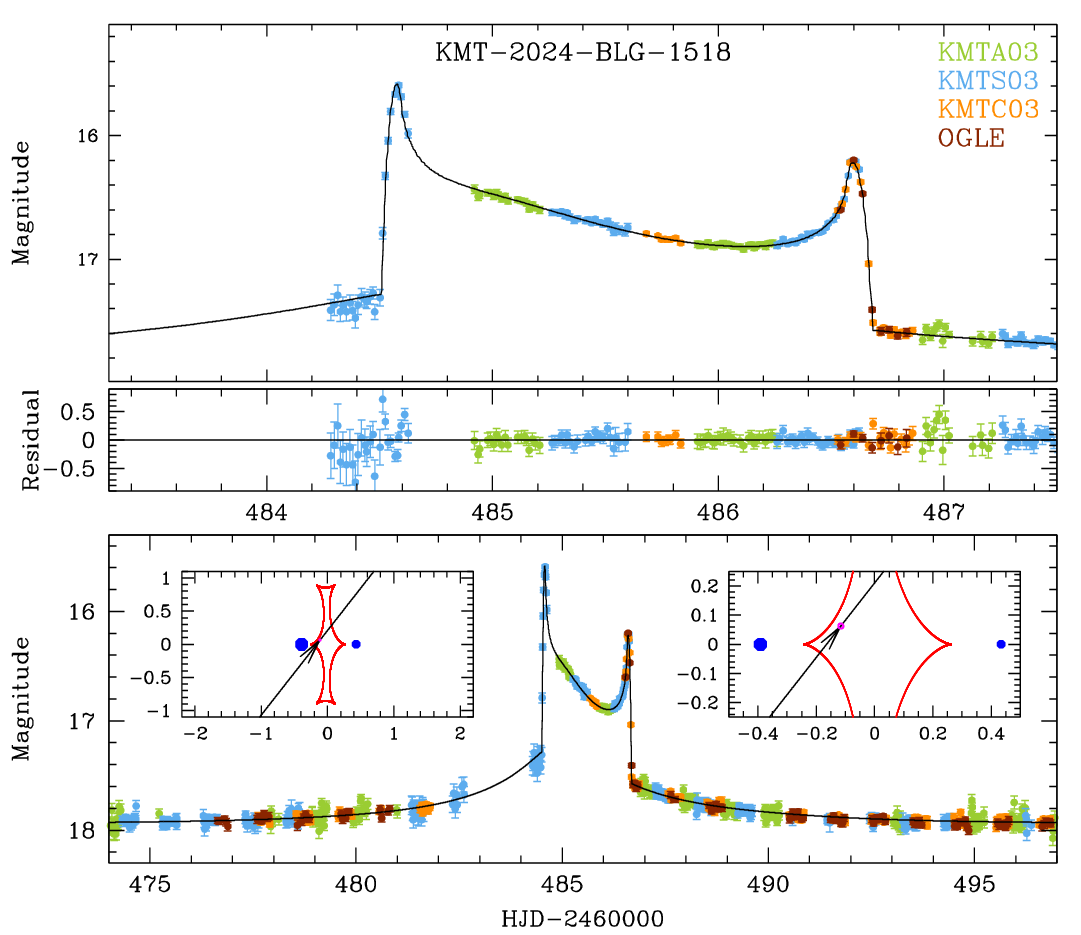}
\caption{
Lensing light curve and lens-system configuration of KMT-2024-BLG-1518.
}
\label{fig:four}
\end{figure}

The insets in the bottom panel of Figure~\ref{fig:three} displays the lens-system 
configuration. Because of the binary separation being close to unity, the lens forms 
a resonant caustic with six cusps, elongated in the direction perpendicular to the 
binary axis. The source traversed the caustic diagonally, entering through the upper 
fold and exiting through the lower-right fold. These fold crossings generated the 
observed caustic spikes. The region between the spikes deviates from a typical U-shaped 
profile because, after entering the caustic, the source asymptotically approached the 
upper-left fold.

\subsection{KMT-2024-BLG-1518} \label{sec:three-four}

The lensing event KMT-2024-BLG-1518, which occurred on a source with a baseline 
magnitude of $I_{\rm base} = 18.0$, was observed by both the KMTNet and OGLE surveys.  
The KMTNet survey initially detected the event on June 25, 2024 (${\rm HJD}^\prime = 
486$), shortly after the light curve exhibited a sharp brightening from the first 
caustic crossing. The OGLE survey identified the event four days later, on June 29, 
2024 (${\rm HJD}^\prime = 490$), when the source experienced a second sharp rise due 
to another caustic crossing.  The source is located in the KMTNet prime field BLG03, 
which mostly overlaps with BLG43, though the source lies in a non-overlapping region. 
As a result, KMTNet observations were conducted with a 30-minute cadence. The $I$-band 
extinction in this field is $A_I = 1.29$.

The lensing light curve of KMT-2024-BLG-1518 is shown in Figure~\ref{fig:four}. It 
features two prominent caustic spikes at ${\rm HJD}^\prime = 484.5$ and 486.6. The 
first caustic crossing was captured by the KMTS data set, while the second was resolved 
through combined observations from OGLE, KMTC, and KMTS. Outside of these spikes, the 
source flux shows a smooth rise and fall, with no additional features.  However, the 
region between the caustic spikes deviates from the typical U-shaped pattern, appearing 
asymmetric with a central portion that exhibits an approximately linear decline.  The 
entire lensing magnification episode was completed within a span of less than 10 days.

Analysis of the light curve reveals that the event is well described by a unique 2L1S 
model. The best-fit binary-lens parameters are $(s, q) \sim (0.82, 0.90)$, indicating 
a binary lens composed of two nearly equal-mass objects with a projected separation 
slightly smaller than the Einstein radius. The full set of lensing parameters is 
provided in Table~\ref{table:three}, and the corresponding model curve is shown in 
Figure~\ref{fig:four}. The event time scale is measured as $\te = 5.118 \pm 0.076$ days, 
with individual time scales of $t_{{\rm E},1} \sim 3.7$ days and $t_{{\rm E},2} \sim 3.5$ 
days for the primary and secondary components, respectively. A detailed examination of the 
caustic-crossing features yields a normalized source radius of $\rho = (8.60 \pm 0.11) 
\times 10^{-3}$. Given that the source is identified as a main-sequence star, this 
relatively large $\rho$ value is unusual for stellar-lens events with similar sources 
and suggests a small angular Einstein radius. The combination of the small $\thetae$ 
and the short event time scale indicates that the lens is likely composed of low-mass 
objects.

The insets in the bottom panel of Figure~\ref{fig:four} illustrate the configuration 
of the lens system.  The source traversed the central region of the caustic, entering 
through the lower-left fold located near the higher-mass lens component and exiting 
through the upper-right fold. As the source traversed the caustic, its trajectory 
closely followed the upper-left fold, resulting in an intra-caustic light curve that 
deviates from the typical U-shaped profile.

\begin{figure}[t]
\includegraphics[width=\columnwidth]{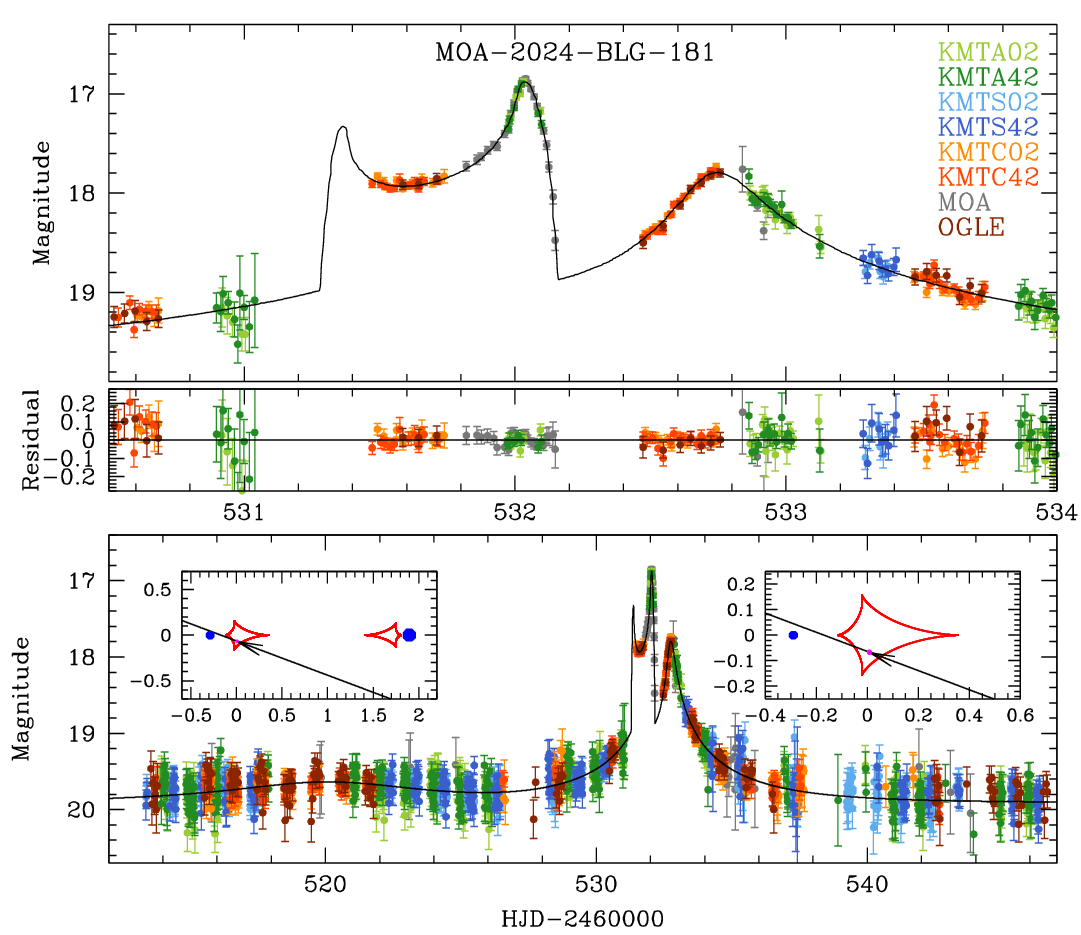}
\caption{
Lensing light curve and lens-system configuration of MOA-2024-BLG-181.
}
\label{fig:five}
\end{figure}

\subsection{MOA-2024-BLG-181} \label{sec:three-five}

The lensing magnification of MOA-2024-BLG-181 was first detected by the MOA group on 
August 9, 2024 (corresponding to ${\rm HJD}^\prime = 531$), during the rising phase 
of the event. This detection was subsequently confirmed by the OGLE and KMTNet surveys. 
The source star was very faint, with a baseline magnitude of $I_{\rm base} = 20.1$, and 
the line of sight toward the field experienced an extinction of $A_I = 1.26$. The event 
occurred in a region covered by both KMTNet prime fields BLG02 and BLG42, enabling 
observations at a high cadence of one data point every 15 minutes.

Figure~\ref{fig:five} presents the lensing light curve of MOA-2024-BLG-181, which 
exhibits three notable features: a caustic spike at ${\rm HJD}^\prime = 532.2$, a 
prominent bump centered around ${\rm HJD}^\prime = 532.8$, and a weaker bump near 
${\rm HJD}^\prime = 520.0$. The spike caused by the source exiting the caustic was 
well resolved in the combined MOA and KMTA data sets, while the caustic entry spike 
was not observed but is estimated to have occurred around ${\rm HJD}^\prime = 531.3$ 
based on an extrapolation of the U-shaped intra-caustic feature. The two bumps are 
likely due to the source passing near caustic cusps. The event was short, with the 
main magnification episode lasting less than 10 days.

Modeling of the light curve yields a unique solution with binary-lens parameters of 
$(s, q) \sim (2.2, 1.7)$. The corresponding model curve is shown in Figure~\ref{fig:five}, 
and the full set of lensing parameters is listed in Table~\ref{table:three}. The event 
time scale is measured as $\te = 6.660 \pm 0.049$ days, indicating a short-duration event. 
The estimated Einstein time scales for the two lens components are approximately 
$t_{{\rm E},1} \sim 4.0$ days for $M_1$ (the component closer to the source trajectory) 
and $t_{{\rm E},2} \sim 5.3$ days for $M_2$ (the farther component). Analysis of the 
caustic-crossing features yields a precise measurement of the normalized source radius 
as $\rho = (6.16 \pm 0.12) \times 10^{-3}$. As with the previous events, this value is 
considerably larger than typical for main-sequence sources, suggesting a small angular 
Einstein radius.

The configuration of the lens system is shown in the insets of the bottom panel of 
Figure~\ref{fig:five}.  Because the binary separation, $s \sim 2.2$, is substantially 
larger than the Einstein radius, the caustic is divided into two segments (composed of 
four folds) located near each of the individual lens components. The source crossed the 
caustic near the lower-mass lens component ($M_1$), producing the observed caustic spikes. 
After exiting the caustic, the source approached the on-axis cusp on the left side, 
resulting in the strong bump observed after the caustic spikes. The bump before the 
spikes was generated as the source passed near the heavier lens component ($M_2$), but 
this bump is weak because the source approached the caustic at a relatively large distance.

\subsection{KMT-2024-BLG-2486} \label{sec:three-six}

The lensing event KMT-2024-BLG-2486 was observed by both the KMTNet and OGLE groups. 
The KMTNet group detected the event at a very early stage on September 9, 2024 
(${\rm HJD}^\prime = 531$), and the OGLE group confirmed the event three days later. 
The source has a baseline magnitude of $I_{\rm base} = 19.6$, and the extinction toward 
the field is $A_I = 2.19$. The source lay in the KMTNet prime fields BLG02 and BLG42, 
and the event was monitored with a 15-minute cadence.

\begin{figure}[t]
\includegraphics[width=\columnwidth]{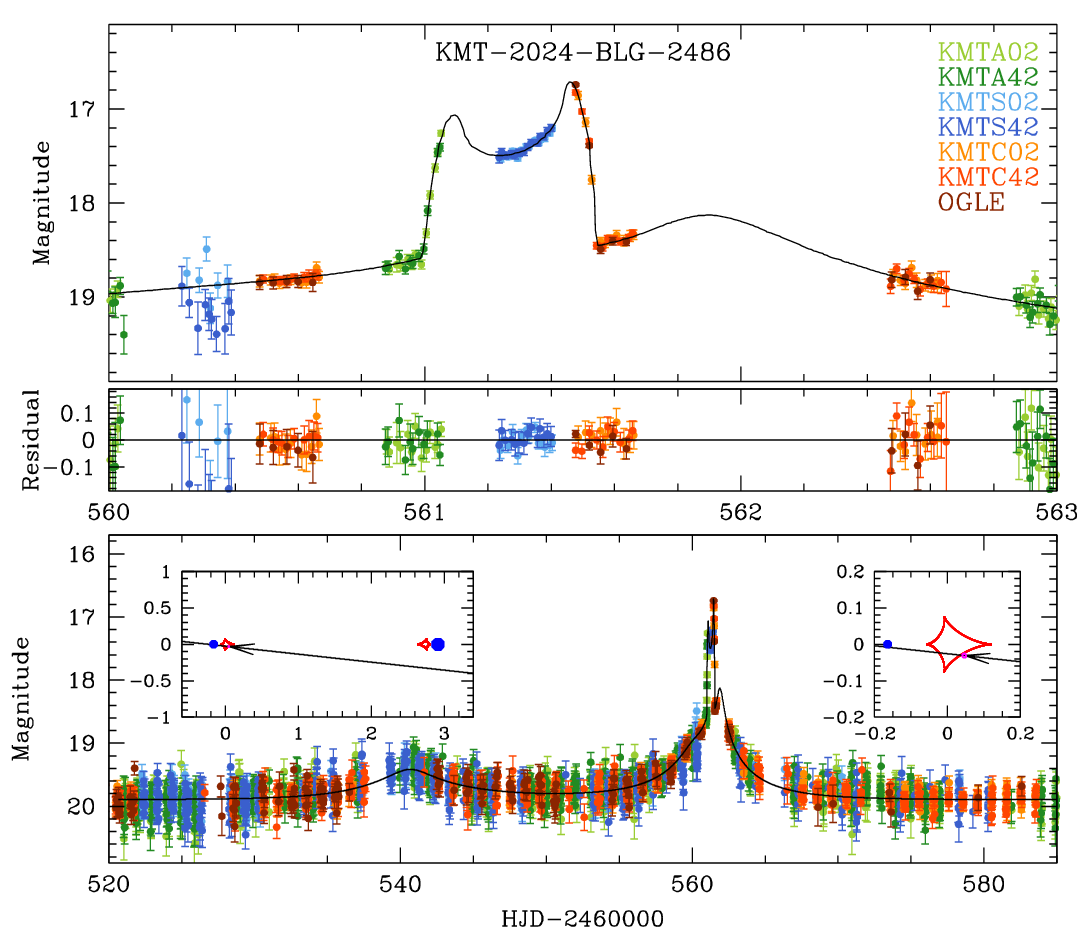}
\caption{
Lensing light curve and lens-system configuration of KMT-2024-BLG-2486.
}
\label{fig:six}
\end{figure}

The lensing light curve of the event is shown in Figure~\ref{fig:six}. It exhibits 
features similar to those of MOA-2024-BLG-181, characterized by a pair of caustic 
spikes at ${\rm HJD}^\prime = 561.1$ and 561.4, with bumps appearing before and after 
the spikes. The pre-spike bump is centered around ${\rm HJD}^\prime = 540.0$, while 
the post-spike bump, although only partially covered, is estimated to be centered 
around ${\rm HJD}^\prime = 562.0$. In addition to the similarity in their anomaly 
features, both events also share the common trait of having short time scales.

\begin{table*}[t]
\small
\caption{Source color, magnitudes, and angular radius.\label{table:four}}
\begin{tabular}{llllllll}
\hline\hline
\multicolumn{1}{c}{Event}                 &
\multicolumn{1}{c}{$(V-I)_S$}             &
\multicolumn{1}{c}{$I_S$}                 &
\multicolumn{1}{c}{$(V-I, I)_{\rm RGC}$}  &
\multicolumn{1}{c}{$I_{RGC,0}$ }         &
\multicolumn{1}{c}{$(V-I)_{s,0}$}         &
\multicolumn{1}{c}{$I_{S,0}$}             &
\multicolumn{1}{c}{$\theta_*$ ($\mu$as)}    \\
\hline
  MOA-2023-BLG-331      &  $2.305 \pm 0.031$  &  $19.868 \pm 0.004$   &  (2.183, 15.792)  &  14.347  & $1.182 \pm 0.050$   & $18.424 \pm 0.020$ &  $1.115 \pm 0.096$  \\
  KMT-2023-BLG-2019     &  $3.583 \pm 0.410$  &  $20.284 \pm 0.004$   &  (3.312, 17.181)  &  14.504  & $1.331 \pm 0.412$   & $17.607 \pm 0.020$ &  $1.798 \pm 0.752$  \\
  KMT-2024-BLG-1005     &  $2.166 \pm 0.238$  &  $19.250 \pm 0.017$   &  (2.166, 15.670)  &  14.339  & $1.060 \pm 0.241$   & $17.920 \pm 0.026$ &  $1.215 \pm 0.305$  \\
  KMT-2024-BLG-1518     &  $1.753 \pm 0.012$  &  $19.209 \pm 0.004$   &  (2.168, 15.704)  &  14.379  & $0.645 \pm 0.042$   & $17.884 \pm 0.020$ &  $0.775 \pm 0.063$  \\
  MOA-2024-BLG-181      &  $2.184 \pm 0.055$  &  $20.177 \pm 0.007$   &  (2.367, 15.919)  &  14.393  & $0.877 \pm 0.068$   & $18.651 \pm 0.021$ &  $0.705 \pm 0.067$  \\
  KMT-2024-BLG-2486     &  $2.551 \pm 0.032$  &  $20.852 \pm 0.008$   &  (2.934, 16.645)  &  14.391  & $0.677 \pm 0.051$   & $18.598 \pm 0.021$ &  $0.577 \pm 0.050$  \\
\hline
\end{tabular}
\tablefoot{$(V - I)_{{\rm RGC},0} = 1.06$.}
\end{table*}

\begin{figure*}[t]
\centering
\includegraphics[width=18.0cm]{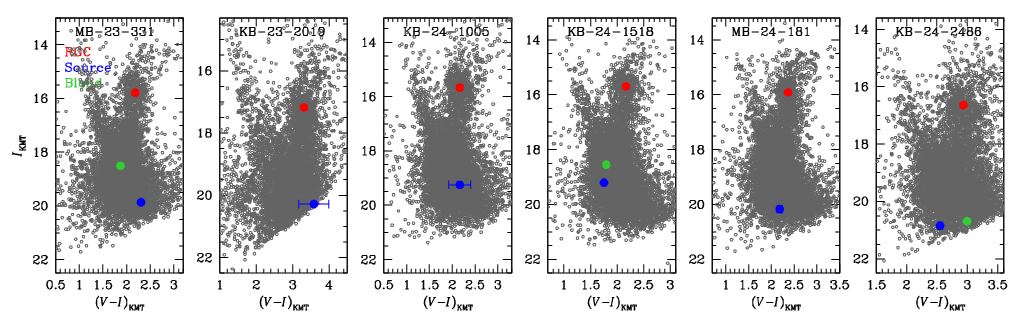}
\caption{
Source positions of events in the instrumental color-magnitude diagrams. In each panel, 
the blue and red dots represent the positions of the source and the centroid of the red 
giant clump (RGC), respectively.  For events with measured blended flux, the locations 
of the blends are also indicated by green dots.
}
\label{fig:seven}
\end{figure*}

Motivated by the similarity of the anomaly pattern to that of MOA-2024-BLG-181, we 
performed a 2L1S modeling analysis, which yielded a unique solution with binary 
parameters $(s, q) \sim (3.0, 1.02)$. This indicates that the lens system comprises 
two nearly equal-mass components separated by approximately three times the Einstein 
radius in projection.  The event time scale was measured to be $\te = 7.395 \pm 0.047$ 
days, which is relatively short. The time scales corresponding to the individual lens 
components are estimated to be $t_{{\rm E},1} \sim 5.2$ days and $t_{{\rm E},2} \sim 
5.3$ days, respectively. In addition, the normalized source radius was determined to 
be $\rho = (5.231 \pm 0.096)\times 10^{-3}$, a value substantially larger than what 
is typically found in events involving M-dwarf lenses with main-sequence source stars. 
This suggests that the Einstein ring of the lens system is relatively small. The model 
curve corresponding to the solution is superposed on the data points in 
Figure~\ref{fig:six}, and the full set of lensing parameters is listed in 
Table~\ref{table:three}.

The configuration of the lens system is depicted in the insets of the bottom panel 
in Figure~\ref{fig:six}.  Consistent with the similarity in the anomaly pattern, 
the lens configuration closely resembles that of MOA-2024-BLG-181. The binary lens 
creates two sets of four-fold caustics near the individual lens components, and the 
source passes through one of these caustics. The source path through the caustic is 
also similar: the source approach to $M_2$ at a relatively large distance produced 
the weak post-spike bump, while the source approach to the on-axis cusp of the caustic 
near $M_1$ led to the stronger post-spike bump.

\section{Source stars and angular Einstein radii} \label{sec:four}

Since the angular Einstein radius is obtained from the normalized source radius using 
the relation 
\begin{equation}
\thetae = { \theta_* \over \rho}, 
\label{eq5}
\end{equation}
estimating $\thetae$ requires knowledge of 
the angular size of the source star. This angular size was determined from the source’s 
color and magnitude after correcting for extinction and reddening. The source color 
and magnitude are also crucial for characterizing the source star itself.

The first step in characterizing a source star involves estimating its instrumental 
color, $(V-I)_S$, and magnitude, $I_S$.  To determine the source magnitudes in the 
$I$ and $V$ bands, we began by generating light curves from the respective $I$- and 
$V$-band data, which were analyzed using the pyDIA photometry code \citep{Albrow2017}.  
The source flux was then estimated by fitting each light curve with a model, $A(t)$, 
according to the relation
\begin{equation}
F_{\rm obs}(t) = A(t) F_S + F_b, 
\label{eq6}
\end{equation}
where $F_{\rm obs}(t)$ is the observed flux at time $t$, and $(F_S, F_b)$ represent 
the fluxes from the source and blended stars, respectively.  Figure~\ref{fig:seven} 
displays the source positions on the instrumental color-magnitude diagrams (CMDs) of 
stars located near the event sources. The CMDs were constructed by performing photometry 
on neighboring stars in a consistent manner using the pyDIA code.

\begin{table}[t]
\caption{Angular Einstein radii.\label{table:five}}
\begin{tabular*}{\columnwidth}{@{\extracolsep{\fill}}lllll}
\hline\hline
\multicolumn{1}{c}{Event}                      &
\multicolumn{1}{c}{$\thetae$ (mas)}            &
\multicolumn{1}{c}{$\mu$ (mas/yr)}            \\
\hline
 MOA-2023-BLG-331     &  $0.192 \pm 0.017  $   &   $10.94 \pm 0.95  $   \\  
 KMT-2023-BLG-2019    &  $0.154 \pm 0.064  $   &   $8.895 \pm 3.722 $   \\  
 KMT-2024-BLG-1005    &  $0.130 \pm 0.033  $   &   $13.257 \pm 3.339$   \\  
 KMT-2024-BLG-1518    &  $0.0901 \pm 0.0074$   &   $6.430 \pm 0.531 $   \\
 MOA-2024-BLG-181     &  $0.115 \pm 0.011  $   &   $6.28 \pm 0.62   $   \\
 KMT-2024-BLG-2486    &  $0.1103 \pm 0.0098$   &   $5.45 \pm 0.48   $   \\
\hline             
\end{tabular*}
\end{table}

In the second step, the source color and magnitude were calibrated by correcting 
for extinction and reddening. This calibration was carried out using a reference 
with well-established de-reddened color and magnitude values.  Specifically, we 
adopted the centroid of the red giant clump (RGC), whose de-reddened color, 
$(V-I)_{{\rm RGC},0}$, and magnitude, $I_{{\rm RGC},0}$, were previously determined 
by \citet{Bensby2013} and \citet{Nataf2013}, respectively.  By measuring the offset, 
$\Delta(V-I, I)$, between the source and the RGC centroid in the instrumental CMD, 
we determined the de-reddened source color and magnitude as
\begin{equation}
(V - I, I)_{s,0} = (V - I, I)_{{\rm RGC},0} + \Delta(V - I, I).
\label{eq7}
\end{equation}
Table~\ref{table:four} presents the instrumental color and magnitude of the source, 
$(V - I, I)_s$, and those of the RGC centroid, $(V - I, I)_{{\rm RGC}}$, along with 
their corresponding de-reddened values, $(V - I, I)_{s,0}$ and $(V - I, I)_{{\rm 
RGC},0}$. The resulting source properties indicate that all sources are main-sequence 
dwarfs: the sources of KMT-2024-BLG-1518 and KMT-2024-BLG-2486 are G-type dwarfs, 
while the others are K-type dwarfs.

In the third step, we estimated the angular radius of the source star using the 
de-reddened color and magnitude derived earlier. Specifically, we applied the $(V-K, 
I)$--$\theta_*$ relation from \citet{Kervella2004}. To use this relation, the measured 
$V-I$ color was first converted to $V-K$ using the color–color transformation of 
\citet{Bessell1988}, and the angular source radius was then inferred from the $(V-K, 
I)$--$\theta_*$ relation. The resulting angular source radii are listed in 
Table~\ref{table:four}.

In the final step, the angular Einstein radius was estimated from the angular source 
radius using the relation given in Eq.~(\ref{eq5}). By combining the estimated angular 
Einstein radius with the event time scale, the relative lens-source proper motion was 
calculated as $\mu = \thetae / \te$. The resulting values of $\thetae$ and $\mu$ for 
the events are presented in Table~\ref{table:five}. For all events, the estimated 
$\thetae$ values are less than 0.2 mas. These small values, together with the short 
time scales, suggest that the lens masses are likely to be low.

\begin{table*}[t]
\caption{Physical lens parameters.\label{table:six}}
\begin{tabular}{llllllll}
\hline\hline                                
\multicolumn{1}{c}{Event}                 &
\multicolumn{1}{c}{$M_1$ ($M_\odot$)}     &
\multicolumn{1}{c}{$M_2$ ($M_\odot$)}     &
\multicolumn{1}{c}{$\dl$ (kpc)}           &
\multicolumn{1}{c}{$a_\perp$ (AU)}        \\ [0.3ex]
\hline
  MOA-2023-BLG-331      &  $0.088^{+0.132}_{-0.050}$  &   $0.137^{+0.206}_{-0.078}$  &  $6.72^{+0.93}_{-1.01}$  &  $1.85^{+0.26}_{-0.28}$  \\  [0.3ex]
  KMT-2023-BLG-2019     &  $0.099^{+0.152}_{-0.057}$  &   $0.033^{+0.051}_{-0.019}$  &  $7.44^{+1.06}_{-1.15}$  &  $1.74^{+0.25}_{-0.27}$  \\  [0.3ex]
  KMT-2024-BLG-1005     &  $0.055^{+0.111}_{-0.029}$  &   $0.020^{+0.040}_{-0.011}$  &  $6.95^{+1.11}_{-1.01}$  &  $0.78^{+0.12}_{-0.11}$  \\  [0.3ex]
  KMT-2024-BLG-1518     &  $0.041^{+0.077}_{-0.022}$  &   $0.037^{+0.069}_{-0.019}$  &  $7.29^{+0.95}_{-0.82}$  &  $0.60^{+0.08}_{-0.07}$  \\  [0.3ex]
  MOA-2024-BLG-181      &  $0.047^{+0.084}_{-0.025}$  &   $0.080^{+0.145}_{-0.044}$  &  $7.60^{+0.94}_{-1.03}$  &  $2.12^{+0.26}_{-0.29}$  \\  [0.3ex]
  KMT-2024-BLG-2486     &  $0.055^{+0.091}_{-0.029}$  &   $0.056^{+0.093}_{-0.030}$  &  $7.61^{+0.97}_{-0.95}$  &  $2.80^{+0.36}_{-0.35}$  \\  [0.3ex]
\hline
\end{tabular}
\end{table*}

\section{Probability of BD pair} \label{sec:five}

We estimated the masses of the lens components to assess the likelihood that 
the lens is a binary composed of BDs.  To accomplish this, we 
conducted a Bayesian analysis that incorporated constraints from the measured 
lensing observables, $\te$ and $\thetae$, along with priors derived from a Galaxy 
model and the mass function of lens objects. The mass function characterizes the 
distribution of lens masses, while the Galaxy model defines the physical and 
dynamical distributions of the lenses.

In the Bayesian analysis, we began by generating a large number of artificial 
events using a Monte Carlo simulation. For each simulated lensing event, the physical 
parameters $(M, \dl, \ds, \mu)_i$ were assigned based on a model mass function and 
Galaxy model.  Specifically, we adopted the mass function from \citet{Jung2022} and 
the Galaxy model from \citet{Jung2021}. The mass function is constructed by adopting 
initial mass function for the bulge population, and the present-day mass function of 
\citet{Chabrier2003} for the disk population.  We considered remnant lenses, such as 
white dwarfs, neutron stars, and black holes, by adopting the \citet{Gould2000b} model. 
In the Galaxy model, the bulge mass distribution is constructed by combining the 
\citet{Han2003} model for the bulge, in which the bulge is modeled as a triaxial 
bar-shaped bulge with parameters derived from infrared star counts, and the modified 
double-exponential model of \citet{Bennett2014} for the disk.  The velocity distribution 
of for bulge stars is constructed based on stars in the Gaia catalog \citep{Gaia2016, 
Gaia2018}, while the distribution of disk stars follows a Gaussian form of \citet{Han1995}, 
and modified to adjust the \citet{Bennett2014} disk model.

Using these parameters obtained from the Monte Carlo simulation, we calculated the 
corresponding lensing observables $(\te, \thetae)_i$ using the relations given in 
Eq.~(\ref{eq1}). Finally, the posteriors for the lens mass and distance were constructed 
by assigning a weight to each artificial event of
\begin{equation}
w_i = \exp \left( -{\chi^2 \over 2} \right);\qquad 
\chi^2 = 
{(t_{{\rm E},i} - \te)^2 \over \sigma(\te)^2} 
+
{(\theta_{{\rm E},i} - \thetae)^2\over \sigma(\thetae)^2}  
\label{eq8}
\end{equation}
Here $(\te, \thetae)$ are the measured values of the lensing observables, and 
$\sigma(\te)$ and $\sigma(\thetae)$ represent their respective uncertainties.

In lensing analyses, among the two observables, $\te$ is directly determined through light
curve modeling, whereas the estimate of $\thetae$ is independently derived from the source’s
physical properties, specifically the angular radius $\theta_*$ inferred from its color and
brightness. Although $\thetae = \theta_*/\rho$, and a correlation between $\te$ and $\rho$
may exist such that $\te$ and $\thetae$ cannot be regarded as completely uncorrelated, in
the present analysis $\thetae$ is obtained through a separate procedure. Therefore, in
practice this correlation is expected to be weak, since $\thetae$ relies on independent 
estimates of the source's angular radius.

\begin{figure}[t]
\includegraphics[width=\columnwidth]{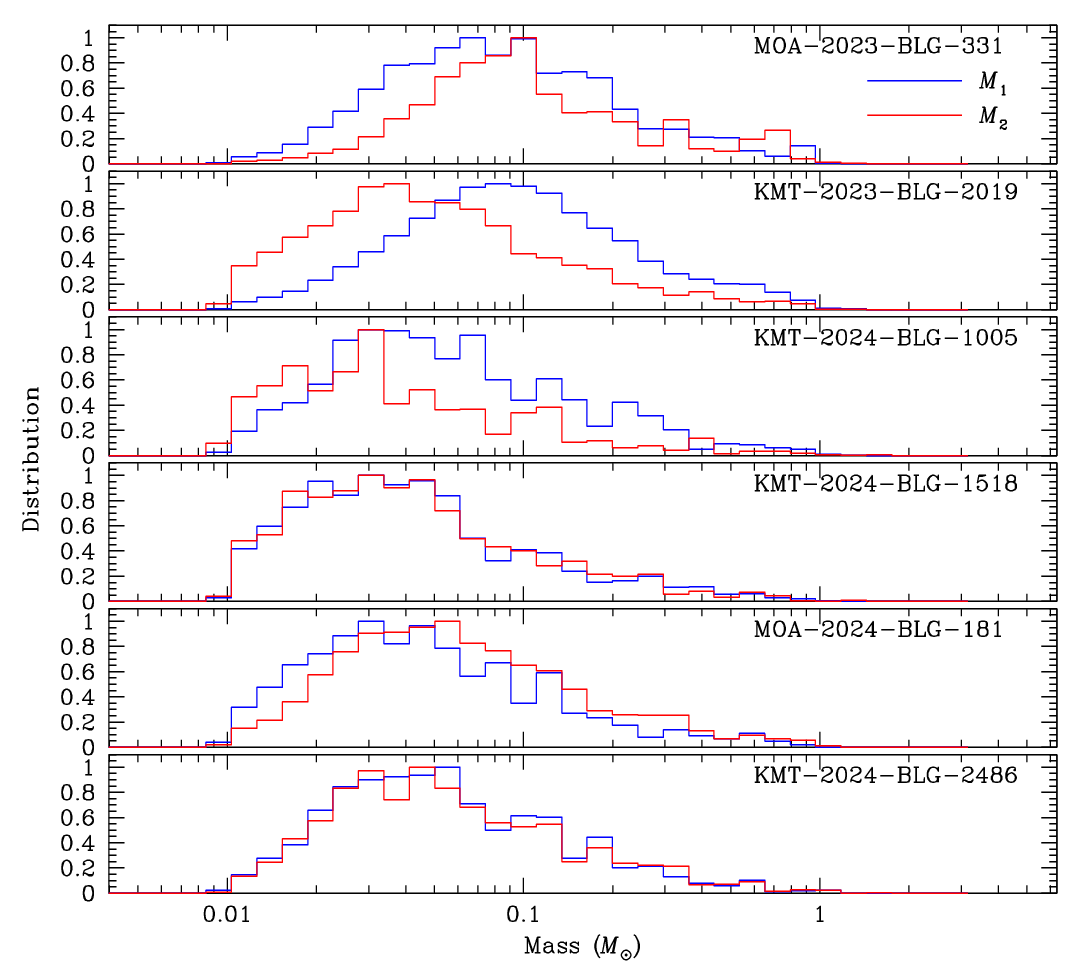}
\caption{
Bayesian posteriors for the lens mass.  The blue and red curves represent the 
posterior distributions of $M_1$ and $M_2$, respectively, where $M_1$ denotes 
the lens component located closer to the source trajectory.
}
\label{fig:eight}
\end{figure}

\begin{figure}[t]
\includegraphics[width=\columnwidth]{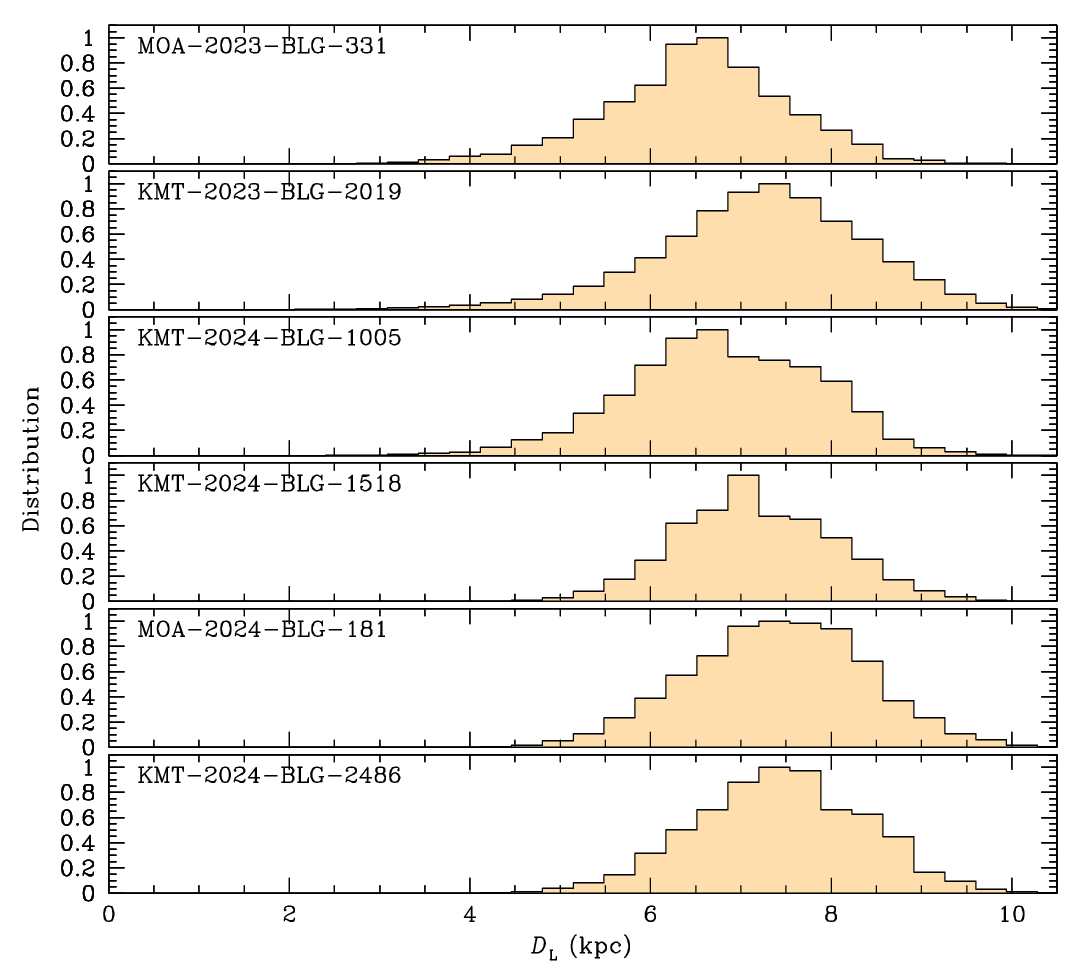}
\caption{
Bayesian posteriors for the distance to the lens.
}
\label{fig:nine}
\end{figure}

Figures~\ref{fig:eight} and \ref{fig:nine} show the posterior distributions for the 
mass and distance of the lenses for each individual event, obtained from the Bayesian 
analyses.  The mass posterior distributions are presented separately for the binary 
lens components $M_1$ and $M_2$, with $M_1$ referring to the lens component that is 
closer to the source trajectory.  Table~\ref{table:six} provides the estimated physical 
parameters of $M_1$, $M_2$, $\dl$, and $a_\perp$, where $a_\perp$ represents the projected 
physical separation between the lens components. For each lens parameter, the median of 
the posterior distribution is chosen as the central value, and the 16\% and 84\% percentiles 
are used to define the lower and upper bounds, respectively.

The values of $p_{\rm BD}$ for $M_1$ and $M_2$ listed in Table~\ref{table:seven} indicate 
the probabilities that the lens components lie within the BD mass range 
of $0.01 \lesssim M/M_\odot \lesssim 0.08$.  For the events KMT-2024-BLG-1005, 
KMT-2024-BLG-1518, MOA-2024-BLG-181, and KMT-2024-BLG-2486, the probabilities that both 
components of the binary lens fall within the BD mass range are greater than 50\%, 
suggesting that these lenses are very likely binary BDs.  In contrast, for MOA-2023-BLG-331L 
and KMT-2023-BLG-2019L, the probabilities that the lower-mass components of the binary 
lenses fall within the BD mass range exceed 50\%, whereas the probabilities for the heavier 
components are below 50\%. This implies that these systems are more likely to consist of 
a low-mass M dwarf and a BD.

It has been determined that the lenses for all events are likely located in the Galactic 
bulge. The values $p_{\rm disk}$ and $p_{\rm bulge}$ in Table~\ref{table:seven} represent 
the probabilities of the lenses being in the disk and bulge, respectively. For all events, 
the probability of the lens being in the bulge ($p_{\rm bulge}$) is significantly higher 
than that of being in the disk ($p_{\rm disk}$). This is also reflected in the posterior 
distributions of $\dl$.

\section{Summary and discussion} \label{sec:six}

With the goal of identifying binary-lens microlensing events in which both lens components 
have substellar masses, we examined binary-lens events detected in the 2023 and 2024 
microlensing survey seasons.  To identify potential binary BD candidates, we applied 
selection criteria that required short event time scales ($\te \lesssim 9$ days) and 
small angular Einstein radii ($\thetae \lesssim 0.17$~mas), both of which are indicative 
of low-mass lenses. An additional requirement was that the event display well-resolved 
caustic features, which are crucial for accurately measuring the angular Einstein radius.  
Through this selection process, we identified six candidate events likely to involve binary 
BD systems: MOA-2023-BLG-331, KMT-2023-BLG-2019, KMT-2024-BLG-1005, KMT-2024-BLG-1518, 
MOA-2024-BLG-181 and KMT-2024-BLG-2486.

\begin{table}[t]
\caption{Probabilities of being in the BD mass range, disk, and bulge.\label{table:seven}}
\begin{tabular*}{\columnwidth}{@{\extracolsep{\fill}}ccccc}
\hline\hline
\multicolumn{1}{c}{Event}            &
\multicolumn{2}{c}{$p_{\rm BD}$ }    &
\multicolumn{1}{c}{$p_{\rm disk}$ }  &
\multicolumn{1}{c}{$p_{\rm bulge}$}  \\
\multicolumn{1}{c}{}                 &
\multicolumn{1}{c}{for $M_1$}        &
\multicolumn{1}{c}{for $M_2$}        &
\multicolumn{1}{c}{}                 &
\multicolumn{1}{c}{}                \\
\hline
  MOA-2023-BLG-331    &  51\%  &  33\%    &   25\%    &  75\%     \\
  KMT-2023-BLG-2019   &  46\%  &  73\%    &   25\%    &  75\%     \\
  KMT-2024-BLG-1005   &  69\%  &  71\%    &   23\%    &  77\%     \\
  KMT-2024-BLG-1518   &  78\%  &  81\%    &   18\%    &  82\%     \\
  MOA-2024-BLG-181    &  75\%  &  54\%    &   30\%    &  70\%     \\
  KMT-2024-BLG-2486   &  70\%  &  70\%    &   17\%    &  83\%     \\
\hline             
\end{tabular*}
\end{table}

Analysis of these events led to models that provided precise estimates for both lensing
observables, $\te$ and $\thetae$. Utilizing the constraints provided by these observables, 
we estimated the masses of the binary components through Bayesian analysis.  The analysis 
indicated that in the cases of KMT-2024-BLG-1005, KMT-2024-BLG-1518, MOA-2024-BLG-181, 
and KMT-2024-BLG-2486, there is a greater than 50\% chance that both components of the 
lens systems have substellar masses, strongly pointing to their nature as binary brown 
dwarfs. In contrast, MOA-2023-BLG-331L and KMT-2023-BLG-2019L suggest mixed-mass systems, 
where the secondary components have a high probability of being brown dwarfs, but the 
primary components are more likely low-mass M dwarfs, indicating these lenses are likely 
composed of an M dwarf–BD pair.

The brown dwarf nature of the binary components can be confirmed through future 
high-resolution imaging with instruments such as the European Extremely Large Telescope.  
If the more massive component ($M_A$) of a binary lens is a star, it will become 
detectable once the source and lens are sufficiently separated. Therefore, if we wait 
until this separation occurs and still do not detect the lens, it would indicate that 
both components are brown dwarfs. Conversely, if the brighter lens component is detected, 
its mass can be measured, which in turn allows us to determine the mass of the secondary 
component and assess whether it falls within the brown dwarf regime.

To estimate the time required to resolve the lens and source, we first compute the 
$K$-band contrast ratio between them as
\begin{equation}
\Delta K = K_{\rm L} - K_S = \Delta K_0 - \Delta A_K = K_{L,0} - K_{S,0} - \Delta A_K
\label{eq9}
\end{equation}
where $\Delta A_K = A_{K,S} - A_{K,L}$ is the difference in extinction between the 
source and lens. For all analyzed events, the lens distances are $\gtrsim 6$~kpc, 
so we adopt $\Delta A_K \approx 0$.  The $K$-band magnitude of the source, 
$K_{{\rm S},0}$, is estimated using the $(V-I, I)_{S,0}$ color and magnitude, together 
with the relation of \citet{Bessell1988}.
The $K$-band magnitude of the lens is estimated as
\begin{equation}
K_{{\rm L},0} = M_{K,{\rm L}} + 5 \log\left( \frac{\dl}{10~{\rm pc}} \right),
\label{eq10}
\end{equation}
where the absolute magnitude is set to $M_{K,{\rm L}} = 10$, corresponding to a 
star with the minimum stellar mass $M_{*,{\rm min}} = 0.08~M_\odot$ (Figure 22 of 
\citealt{Benedict2016}) which is the hydrogen-burning limit at the boundary between 
stars and brown dwarfs.  The distance to the lens is given by $ \dl = {\rm AU}/
(\pi_{\rm rel} + \pi_{\rm S})~{\rm kpc}$, where $\pi_{\rm rel} = \theta_{{\rm E},A}^2/
(8.14~M_{*,{\rm min}})$ is the lens-source relative parallax (in mas), and $\pi_{\rm S} 
= {\rm AU}/(8.5~{\rm kpc}) \approx 0.117~{\rm mas}$ is the parallax of the source, 
assuming $\ds = 8.5$~kpc.  The angular Einstein radius corresponding to the primary 
mass $M_A = M_{*,{\rm min}}$ is $\theta_{{\rm E},A}= [1/(1+q)]^{1/2} \thetae$.

\begin{table}[t]
\caption{Contrast ratio between the lens and source.\label{table:eight}}
\begin{tabular*}{\columnwidth}{@{\extracolsep{\fill}}ccccc}
\hline\hline
\multicolumn{1}{c}{Event}           &
\multicolumn{1}{c}{$K_{S,0}$ }      &
\multicolumn{1}{c}{$K_{L,0}$ }      &
\multicolumn{1}{c}{$\Delta K_0 $}   \\
\hline
  MOA-2023-BLG-331    &   16.88     &   24.1   &   7.2    \\
  KMT-2023-BLG-2019   &   15.99     &   24.2   &   8.2    \\
  KMT-2024-BLG-1005   &   16.52     &   24.3   &   7.8    \\
  KMT-2024-BLG-1518   &   17.14     &   24.5   &   7.4    \\
  MOA-2024-BLG-181    &   17.59     &   24.4   &   6.8    \\
  KMT-2024-BLG-2486   &   17.82     &   24.5   &   6.7    \\
\hline             
\end{tabular*}
\end{table}

Table~\ref{table:eight} lists the contrast ratios for the individual events, with 
$\Delta K_0$ values ranging from 6.7 to 8.2, corresponding to flux ratios between 500 
and 2000. Due to this high contrast, resolving the lens from the source would require 
a separation of approximately 8–10 times the FWHM, which translates to about 110--140 
mas. Assuming the lower limit of 110 mas, the lens and source could be resolved after 
a wait time of approximately 8 years (in 2032) for KMT-2024-BLG-1005, which has the 
highest proper motion of $\mu \sim 13.3$~mas/yr.  For KMT-2024-BLG-2486, which has the 
lowest proper motion of $\mu \sim 5.5$~mas/yr, the required separation would be reached 
in about 20 years (in 2044).

Given that six candidates were identified in just two years of data, it is likely that 
additional brown dwarf binaries can be discovered in the full microlensing survey dataset.
Furthermore, more precise mass measurements could, in principle, be obtained by measuring 
microlens parallax in events observed by future space-based missions such as the Nancy 
Grace Roman Space Telescope and the Chinese Space Station Telescope.

\begin{acknowledgements}
This research was supported by the Korea Astronomy and Space Science Institute under 
the R\&D program (Project No. 2025-1-830-05) supervised by the Ministry of Science and 
ICT.  This research has made use of the KMTNet system operated by the Korea Astronomy 
and Space Science Institute (KASI) at three host sites of CTIO in Chile, SAAO in South 
Africa, and SSO in Australia.  Data transfer from the host site to KASI was supported 
by the Korea Research Environment Open NETwork (KREONET).  C.Han acknowledge the support 
from the Korea Astronomy and Space Science Institute under the R\&D program (Project No. 
2025-1-830-05) supervised by the Ministry of Science and ICT.  J.C.Y., I.G.S., and S.J.C. 
acknowledge support from NSF Grant No. AST-2108414.  W.Zang acknowledges the support from 
the Harvard-Smithsonian Center for Astrophysics through the CfA Fellowship.  The MOA 
project is supported by JSPS KAKENHI Grant Number JP16H06287, JP22H00153 and 23KK0060.
C.R. was supported by the Research fellowship of the Alexander von Humboldt Foundation.
\end{acknowledgements}

\end{document}